\documentclass[aps,pre,twocolumn,a4paper,amsfronts,showpacs]{revtex4}
\usepackage{amssymb,amsmath,amsthm}
\usepackage{graphics}
\usepackage{epsfig}
\begin{document}
\title{ One-way coupled  Van der Pol system}
\author{Sangeeta Ghosh and B. Talukdar}
\affiliation{ Department of Physics, Visva-Bharati University, Santiniketan, 731235, India}
 \author{U. Das}
\affiliation{Department of Physics, Abhedananda Mahavidyalaya, Sainthia, 731234, India}
\begin{abstract}
The equation of the Van der Pol oscillator, being characterized by a dissipative term, is non-Lagrangian. Appending an additional degree of freedom we bring the equation in the frame of action principle and thus introduce a one-way coupled system. As with the Van der Pol oscillator, the coupled system also involves only one parameter that controls the dynamics. The response system is described by a linear differential equation coupled nonlinearly to the drive system. In the linear approximation the equations of our coupled system coincide with those of the Bateman dual system (a pair of damped and anti-damped harmonic oscillators). The critical point of damped and anti-damped oscillators are stable and unstable for all physical values of the frictional coefficient $\mu$. Contrarily, the critical points of the drive- (Van der Pol) and response systems depend crucially on the values of $\mu$. These points are unstable for $\mu > 0$ while the critical point of the drive system is stable and that of the response system is unstable for $\mu < 0$. The one-way coupled system exhibits bifurcations which are different from those of the uncoupled Van der Pol oscillator. Our system is chaotic and we observe phase synchronization in the regime of dynamic chaos only for small values of $\mu$.
\end{abstract}

\maketitle
\section{Introduction}
A great variety of periodic processes, ranging from reoccurrence of economic crisis to beating of the human heart, takes place in such a way that external energy is supplied to the system over a part of the period and dissipated within the system in another part of the period. Periodic processes of this type have been qualified as relaxation oscillations. In fact, the relaxation oscillator constitutes a self-sustained system with a built in mechanism to pump energy to oscillations that are too small and damp those oscillations that grow too large. It appears that Van der Pol \cite{1} was the first to model relaxation oscillations in a laboratory experiment by considering an electric circuit which consists of a battery connected in series with a resistor and a capacitor. A diode valve which acts as a nonlinear device was also connected across the capacitor.\par Because of their applicative relevance, the driven and coupled Van der Pol oscillators have been extensively studied in the literature \cite{2}. These oscillators display physical and mathematical properties which are radically different from those of an isolated Van der Pol oscillator.   In the present work we are interested to construct a model for the one-way coupled system \cite{3} in which Van der Pol oscillator is a drive. Initiated by Chua \cite{4}, studies in the dynamical properties of one-way coupled systems have become a subject of considerable interest because such systems exhibit bifurcation, chaos and synchronization. Thus we can expect that studies in our newly proposed system may also lead to similar physically significant results. \par In section 2 we note that Van der Pol oscillator is not invariant under time reversal and thus represents a physically incomplete system \cite{5} in the sense of the Hamilton's variational principle. We follow Bateman \cite{6} and double the degree of freedom to bring in an associated equation such that the system of equations follows from the action principle. Curiously enough, this automatically provides us with a one-way coupled system in which Van der Pol oscillator is a drive. In section 3 we write the equations for coupled Van der Pol system as a system of first order differential equations that provide a basis to study the associated dynamics. For an uncoupled Van der Pol oscillator the stability of critical points is well known. It has a limit cycle attractor and bifurcation structure dependent on the parameter of the oscillator. We shall study the stability of our one-way coupled system with particular emphasis on the nature of the attractor and bifurcation that lead to chaos. One of the remarkable features of coupled nonlinear oscillators is synchronization which characterizes the emergence of coherent motion among the constituent oscillators. Keeping this in view we shall also analyze the stability and optimization of the synchronization process which is likely to occur in our system. Finally, in section 4 we shall make some concluding remarks.
\section{One-way coupled system : Action principle and linear analog}
 The equation of the Van der Pol oscillator is given by \cite{1}
\begin{equation}
 G(\ddot x,\dot x,x) = \ddot x-\mu(1-x^2)\dot x+x = 0, 
\end{equation}

where overdots denote differentiation with respect to time and $\mu$ is a parameter of the system. Clearly, this oscillator is driven by a velocity dependent force and provides a typical example of the dissipative system.  Moreover, the equation violates the time reversal symmetry. Such equations do not follow from the action principle to have a Lagrangian representation. It is well known that a dynamical equation can be derived by using the Hamilton's variational principle if it satisfies the criterion \cite{7}
\begin{equation}
\frac{\partial G(\ddot x,\dot x,x)}{\partial\dot x} = \frac{d}{dt}\frac{\partial G(\ddot x,\dot x,x)}{\partial\ddot x}.
\end{equation}
It is easy to see that $(1)$ invalidates the condition in $(2)$ such that the equation of the Van der Pol oscillator is non-Lagrangian. A non-Lagrangian system can be brought within the framework of variational principle by doubling the degrees of freedom of the equation of motion \cite{6}. To illustrate this consider the time reversal symmetry breaking equation of the linearly damped harmonic oscillator
\begin{equation}
\ddot y +\mu \dot y + y = 0 .
\end{equation}
The quantity $\mu$ now stands for the dissipation constant of the medium in which the oscillator is embedded.
Let us write a Lagrangian
\begin{equation}
L = x(\ddot y+\mu \dot y + y),
\end{equation}
characterized by two degrees of freedom, namely $x$ and $y$. Clearly, the Euler-Lagrange equation in $x$ gives $(3)$.  To obtain the equation of motion for the $x$ variable consider the Euler-Lagrange equation \cite{8}
\begin{equation}
\frac{d^2}{dt^2}\left(\frac{\partial L}{\partial \ddot y}\right) - \frac{d}{dt}\left(\frac{\partial L}{\partial\dot y}\right) + \frac{\partial L}{\partial y} = 0,
\end{equation}
in the $y$ variable. From $(4)$ and $(5)$ we get
\begin{equation}
\ddot x - \mu \dot x + x = 0.
\end{equation}
Equation $(6)$ models a physical system which absorbs energy dissipated in the damped harmonic oscillator $(3)$.  The system represented by $(3)$ and $(6)$ is often called dual system of Bateman who initiated the idea of bringing non-Lagrangian systems within the frame of action principle. However, it is not immediately clear how the transfer of energy takes place between the components of the system. In the following we show that the approach of Bateman when applied to the equation of a Van der Pol oscillator we get a physically transparent one-way coupled system with the Van der Pol oscillator as the drive system.\par To bring the equation of motion $(1)$ in the frame of the action principle let us define a Lagrangian
\begin{equation}
L = y(\ddot x - \mu(1-x^2)\dot x + x),
\end{equation}
so as to write an associated equation
\begin{equation}
\ddot y + \mu(1-x^2)\dot y + y = 0,
\end{equation}
similar to that in $(6)$. As opposed to the uncoupled equations $(3)$ and $(6)$ , we have now a coupled system in which the Van der Pol oscillator drives the associated response system through nonlinear coupling. All contemporary physical theories use Hamilton's variational principle to decode the laws of nature and the one-way coupled system obtained by us follows from the constraint imposed by this principle. Looking from this point of view the equations in $(1)$ and $(8)$ which represent a one-way coupled system appear to be a natural selection. This is not true for other widely cited examples of one-way coupled systems. In this context we note that the response system in $(8)$ is governed by a linear differential equation coupled nonlinearly to the Van der Pol oscillator. Interestingly, if this coupling were absent and if we would linearize the equation of the Van der Pol osillator, we will get back the equation for the Bateman dual system. Thus the Bateman system can be considered as a linear analog of our one-way coupled system.
\section{Stability Analysis}
For stability analysis and even for purposes of numerical simulations, it is convenient to write $(1)$ and $(8)$ as a system of first order differential equations
\begin{subequations}
\begin{equation}
\dot x_1 = x_2,
\end{equation}
\begin{equation}
\dot x_2 = \mu(1-x_1^2)x_2 - x_1,
\end{equation}
\end{subequations}
\begin{subequations}
\begin{equation}
\dot y_1 = y_2,
\end{equation}
and
\begin{equation}
\dot y_2 = -\mu(1-x_1^2)y_2 - y_1.
\end{equation}
\end{subequations}
In writing $(9)$ and $(10)$, we have used $x_1$ for $x$ and $y_1$ for $y$. Clearly, $x_1 = x_2 = y_1 = y_2 = 0$ represents the fixed point of $(9)$ and $(10)$. Linearizing the system about the fixed point , say, $\zeta$ and $\eta$ we have
\begin{subequations}
\begin{equation}
\dot \zeta_1 = \zeta_2,
\end{equation}
\begin{equation}
\dot \zeta_2 = \mu \zeta_2 - \zeta_1,
\end{equation}
\begin{equation}
\dot \eta_1 = \eta_2,
\end{equation}
and 
\begin{equation}
\dot \eta_2 = -\mu \eta_2 - \eta_1.
\end{equation}
\end{subequations}
In writing $(11)$, we have used the same notational convention as that adopted for $x$ and $y$.
The Jacobian matrix for $(11)$ given by

 $$ \,\,\,\,\,\,J = \left(\begin{array}{cccc}
         0 & 1 & 0 & 0\\
	 -1 & \mu & 0 & 0\\
	 0 & 0 & 0 & 1\\
	0 & 0 & -1 & -\mu
\end{array}\right)\eqno(12)$$
has the eigenvalues\\

\begin{subequations}
\begin{equation}
\lambda_1=\frac{\mu}{2}+\frac{1}{2}\sqrt{\mu^2-4},
\end{equation}
\begin{equation}
\lambda_2=\frac{\mu}{2} - \frac{1}{2}\sqrt{\mu^2-4}, 
\end{equation}
\begin{equation}
\lambda_3=-\frac{\mu}{2} + \frac{1}{2}\sqrt{\mu^2-4},
\end{equation}
and 
\begin{equation}
\lambda_4=-\frac{\mu}{2} -\frac{1}{2}\sqrt{\mu^2-4}.
\end{equation}
\end{subequations}
The nature of the eigenvalues gives a broad classification scheme for the equilibrium points regarding their stabily and instability. We shall use the method for such classification wherever necessary.
\par 
If the nonlinear terms in $(1)$ and $(8)$ were absent the drive and response systems will go over to the Bateman equations $(6)$ and $(3)$. Thus it will be interesting to compare the phase diagrams of $(1)$ and $(6)$ and of $(8)$ and $(3)$ in order to study the points of contrast or of similarity for the phase-space evolution of the one-way coupled system and Bateman dual system. 
\par 
To draw the phase diagrams of Bateman system we rewrite $(3)$ and $(6)$ as a system of first order differential equation
\begin{subequations}
\begin{equation}
 \dot y_1=y_2,
\end{equation}
\begin{equation}
\dot y_2=-\mu y_2 - y_1,
\end{equation}
\end{subequations}
and 
\begin{subequations}
\begin{equation}
\dot x_1=x_2,
\end{equation}
\begin{equation}
\dot x_2=\mu x_2 -x_1.
\end{equation}
\end{subequations}
\begin{figure}[h]
\begin{center}
\includegraphics[width=5cm]{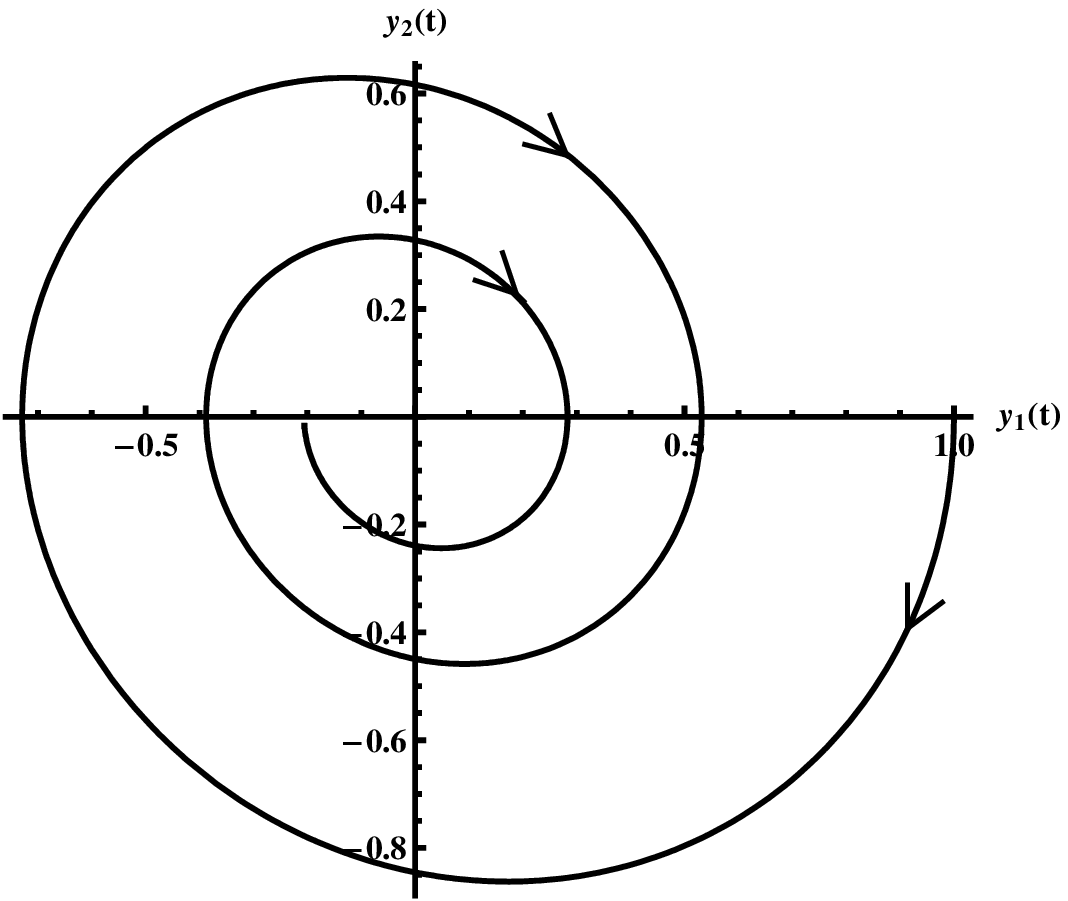}
\includegraphics[width=5cm]{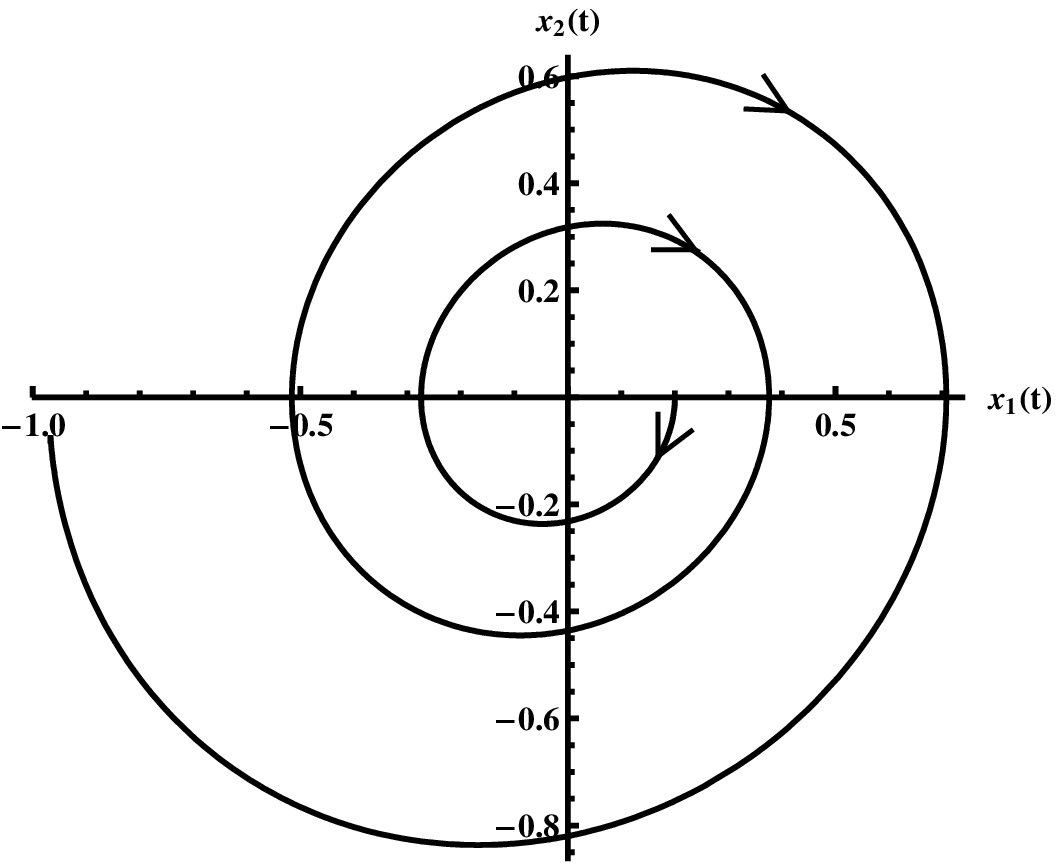}
\caption{Phase diagrams of Bateman dual system.$(a)$ Damped harmonic oscillator and $(b)$ The associated system.}
\end{center}
\end{figure}
\begin{figure}[h]
\begin{center}
\includegraphics[width=5cm]{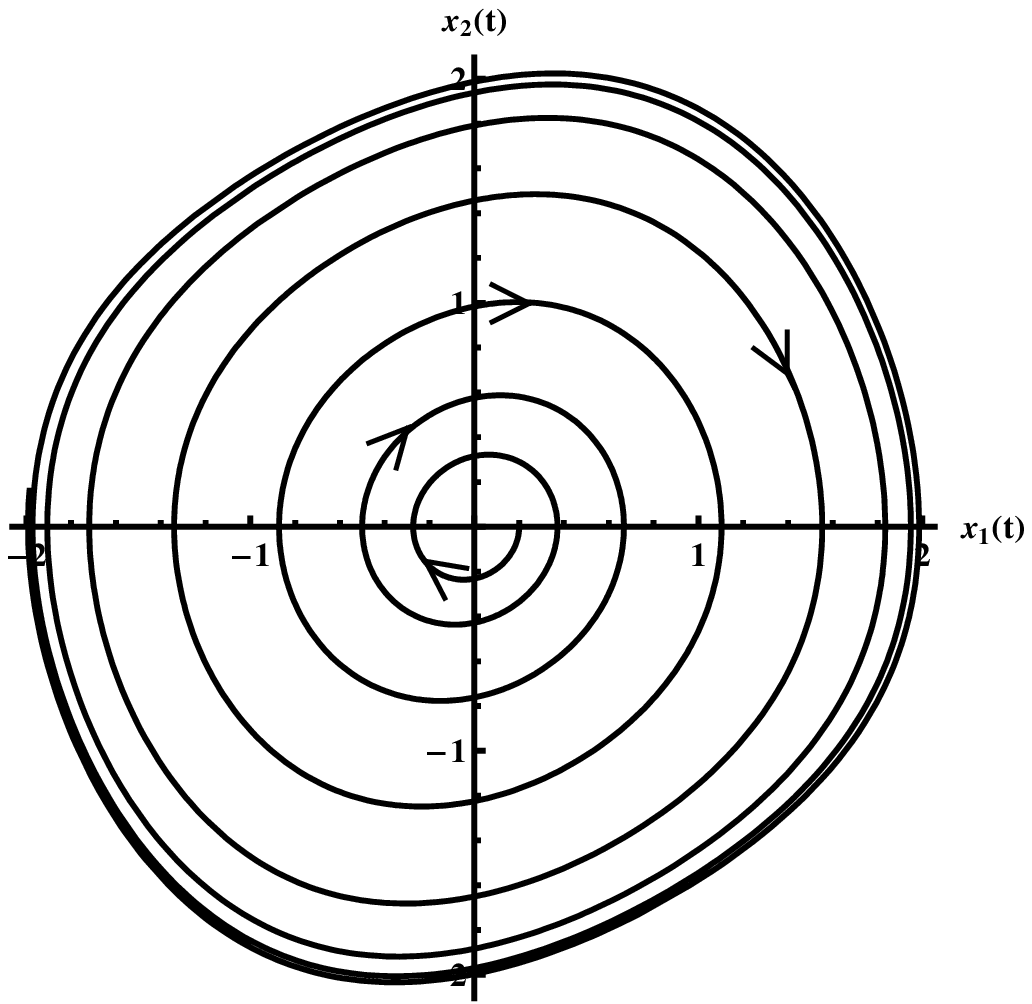}
\includegraphics[width=5cm]{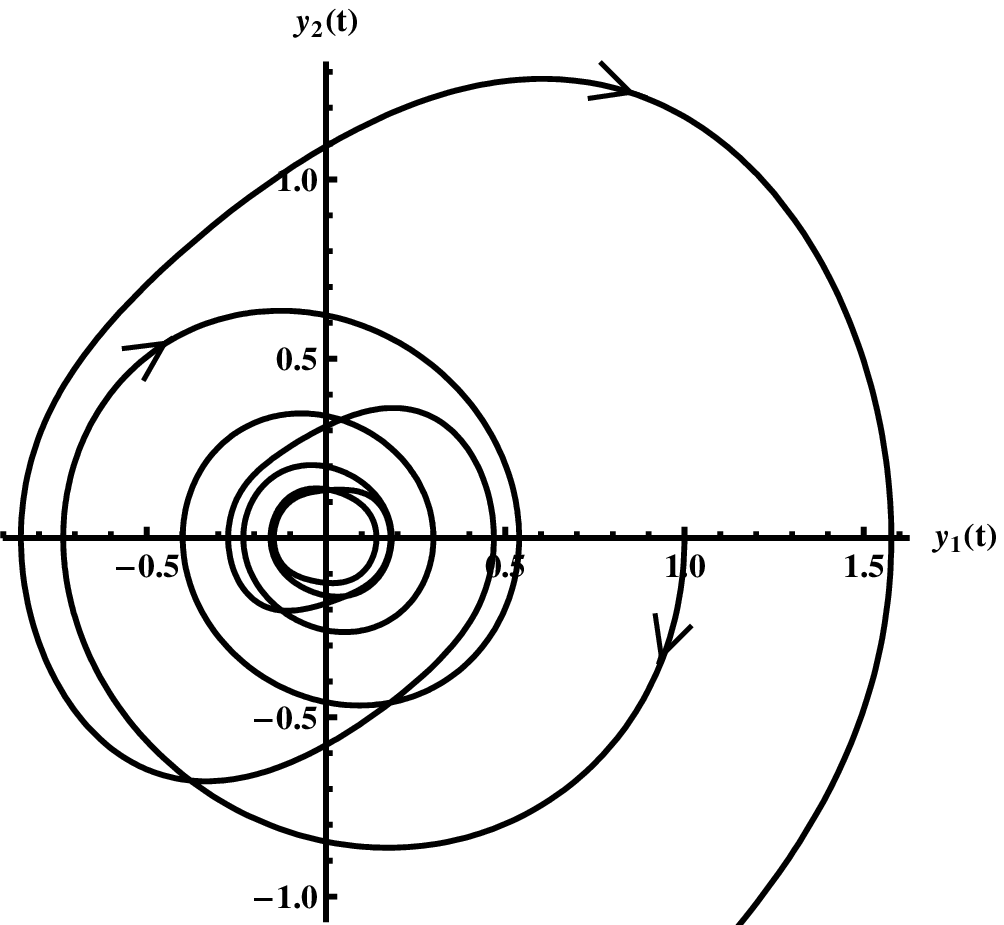}
\caption{Phase diagrams of the one-way coupled system. $(a)$ Drive system (Van der Pol oscillator) and $(b)$ Response system (Equation $(8)$)}
\end{center}
\end{figure}
As with $(9)$ and $(10)$, we used $y_1 = y$ and $x_1 = x$ to write the equations in $(14)$ and $(15)$. We solved this set of equations using the initial conditions $y_1(0) = 1$, $y_2(0) = 0$ and $x_1(0) = 0.2$, $x_2(0) = 0$. We have chosen to work with $\mu = 0.2$. The phase diagrams of the damped harmonic oscillator $(3)$ is given in Fig.$1(a)$. Here we see that the phase trajectory starting from the point $(1,0)$ spirals in and tends to meet the origin such that the fixed point is a stable focus.We have presented in Fig.$1(b)$ a similar phase diagram for the associated equation in $(6)$. The phase trajectory starts from $(0.2,0)$ and spirals out. Thus the fixed point is an unstable focus. We shall now make similar comparison between the phase diagrams of $(1)$ and $(8)$.We solved the system of equations $(9)$ and $(10)$ with the same initial conditions as used for $(13)$ and $(14)$. Fig.$2(a)$ gives the phase diagram for the Van der Pol oscillator. As expected the phase trajectory shows a stable limit cycle with origin as an unstable focus for $\mu = 0.2$. The phase diagram (Fig.$2(b)$) of the response system, which in conjunction with the Van der Pol oscillator forms a one-way coupled system, exihibits some interesting features. For example, starting from $(1,0)$ the phase trajectory tends to form a small amplitude limit cycle. Then all of a sudden it spirals out. \par Since in the absence of the nonlinear term, $(1)$ concides with $(6)$, we will compare the phase trajectory in Fig.$2(a)$ with that in Fig.$1(b)$. Looking closely into these figures we see that the phase trajectory in Fig.$1(b)$ spirals out without forming a limit cycle while the phase trajectory in Fig.$2(a)$ has a stable limit cycle. Thus the formation of limit cycle is a typical feature of sustained oscillation. Again comparison between the phase trajectories of Fig.$2(b)$ and Fig.$1(a)$ leads us to more surprising points of contrast because the phase trajectory in Fig.$1(a)$ simply spirals in rather than showing any complicated behavior as found for the phase trajectory in Fig.$2(b)$. Ideally, one would expect some similarities between the phase trajectories of Fig.$2(b)$ and Fig.$1(a)$.This is, however, not true. The phase-space evolution of the drive system is radically different from that of the damped harmonic oscillator because the energy transfer between the systems represented by $(1)$ and $(8)$ is explicit while the equations of the Bateman dual system are uncoupled.
\par
\begin{figure}[h]
\begin{center}
\includegraphics[width=6cm]{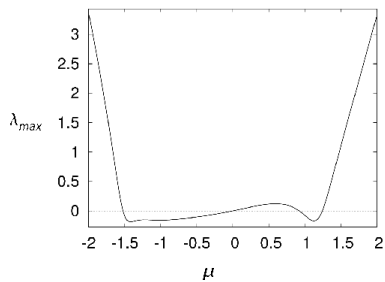}
\caption{Maximum Lyapunov exponent versus parameter of the system}
\end{center}
\end{figure}
We have drawn the phase diagrams of our one-way coupled system for $\mu = 0.2$ which satisfies the inequality $0 < \mu < 2$. If $\mu$ is constrained by this inequality the uncoupled Van der Pol oscillator will not exhibit any bifurcation. However, there are three critical values of $\mu$ for which certain qualitative changes, namely, bifurcations occur in the dynamics of $(1)$. For example at $\mu = -2$ there occurs a bifurcation in which the critical point changes from a stable node to a stable focus. The radicals in $\lambda_1$ and $\lambda_2$ (eigenvalues of the equation for the uncoupled Van der Pol oscillator) do not depend on the sign of $\mu$. When the real parts of the eigenvalues go through zero ($\mu = 0$) with non-zero complex parts we have Hopf bifurcation. At $\mu = 2$ a bifurcation from an unstable focus to an unstable node occurs.\par Consider a similar situation for the coupled system. The response system represented by $(8)$ has also three bifurcation points. For $\mu > -2$ the phase trajectory diverges from the critical point such that there is no attractor in the neighborhood of the equilibrium point. As we increase the value of $\mu$ and choose to work in the region $-2 < \mu < 0$ the equilibrium point becomes an unstable focus similar to that in Fig.$1(b)$. Thus at $\mu = -2$, a bifurcation takes place in which a diverged state changes to an unstable focus. If we still increase the value of $\mu$ through zero $(0 < \mu <2)$ there appears a fold limit cycle (Fig.$2(b)$). This small amplitude limit cycle is neither stable nor unstable. As noted earlier the phase trajectory first spirals in, tends to form a limit cycle, then all of a sudden spirals out. This implies that at $\mu = 0$ a bifurcation occurs from a unstable focus to a fold limit cycle. For $\mu \geq 2$ the fold limit cycle disappears and the phase trajectory again diverges from its critical point. So at $\mu = 2$, a bifurcation occurs in which a fold limit cycle vanishes and the state of the system diverges.\par In the above we have seen that both uncoupled Van der Pol oscillator and response system driven by it exhibit bifurcations at $\mu = -2$, $0$ and $2$ although the type of bifurcation is different in each case. Since the appearance of bifurcation in a dynamical system signals onset of chaos, it will be interesting to examine which of these system is chaotic for $ -2 \leq \mu  \leq 2$.
\begin{figure}[h]
\begin{center}
\includegraphics[width=6cm]{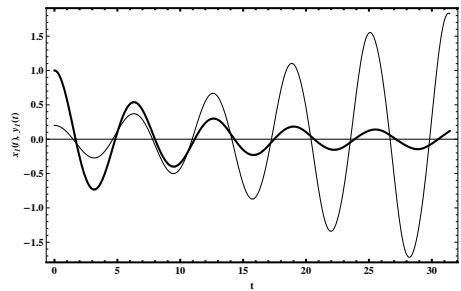}
\caption{ $x(t)$ (thin line) and $y(t)$ (thick line) as a function of $t$}
\end{center}
\end{figure}
\par 
The non-wandering set is a set of points in the phase space for which all orbits starting from a point of the set come arbitrarily close and arbitrarily often to any point in the set. The different types of non-wandering sets include fixed point, limit cycle, quasi-periodic orbits and chaotic orbits. A non-wandering set can be either stable/unstable as one changes the parameter of the system. In particular, such a set is said to be Lyapunov stable if every orbit starting in its neighborhood remains in its neighborhood else it is unstable. The linear stability analysis of the system leads, in a rather natural way, to the concept of Lyapunov stability. Let $X(t)$ be an orbit satisfying 
\begin{equation}
 \dot X(t)= f(X,t) .
\end{equation}
A solution $X(t)$ of $(15)$ will be asymptotically stable if any infinitesimally small perturbation $\delta X(t)$ decays. In order to determine the stability of any solution $X_0(t)$ of $(15)$ assume 
\begin{equation}
 X(t) =X_0(t) + \delta X(t),
\end{equation}
as one of the purturbed solution of $(15)$. Using $(16)$ in $(15)$ we expand the function $f(X,t)$ in Taylor series to get
\begin{equation}
 \frac{d}{dt} \Delta X = \frac{df}{dX}\Delta X .
\end{equation}
The linearized equation of motion for $\Delta X$ in $(17)$ is justified as long as the orbit is in the neighborhood of $X_0$. At the fixed point $\frac{df}{dX}$ = a constant (say  $\lambda)$. Thus 
\begin{equation}
 \Delta X = e^{\lambda_s t} \Delta X_s,
\end{equation}
where, $\lambda_s$ and $X_s$ are eigenvalue and eigenvector of the Jacobian matrix of the linearized system and $s$ is the dimension of the phase space. There are as many Lyapunov exponents as dimensions of the phase space.\par  The Lyapunov exponent $\lambda_s$ can be calculated for each dimension. When talking about a single exponent one is normally referring to the largest. If the Lyapunov exponent is positive the system is chaotic and unstable. In this case the nearby points will diverge irrespective of how close they are. Thus the magnitude of the Lyapunov exponent is a measure of chaosness in the system. 
\begin{table}[h]
\caption{Signs of $\lambda_{max}$ in different ranges of the parameter $\mu$}
\begin{tabular}{|c|c|c|c|}
\hline
Sl. No. & Range of $\mu$ & Sign of $\lambda_{max}$ & Nature of the system\\
\hline
$1.$ & $\mu < -1.5$ & positive & chaotic\\
$2.$ & $\mu = -1.5$ & zero & conservative\\
$3.$ & $-1.5< \mu < 0$ & negative & dissipative\\
$4.$ & $\mu = 0$ & zero & conservative\\
$5.$ & $0<\mu < 0.92$ & positive & chaotic\\
$6.$ & $\mu = 0.92$ & zero & conservative\\
$7.$ & $0.92<\mu < 1.25$ & negative & dissipative\\
$8.$ & $\mu = 1.25$ & zero & conservative\\
$9.$ & $\mu > 1.25$ & positive & chaotic\\
\hline
\end{tabular}
\end{table}
\par
In Fig.$3$ we plot the Lyapunov exponent, $\lambda_{max}$, versus $\mu$, the only parameter of our system. The curve for $\lambda_{max}$ clearly shows that the exponent takes up positive, zero and negative values such that the corresponding motion of the system is chaotic, neutrally stable (conservative) and dissipative. Table 1 gives the signs of $\lambda_{max}$ in different ranges of the parameter. This helps us identify the nature of the system as a function of $\mu$.
Fig.$2(b)$ gives the phase diagram of the response system for $\mu = 0.2$. This $\mu$ value satisfies the inequality $0 < \mu < 0.92 $(Sl. No.$5$). Thus the curve in this figure gives the phase space evolution of the system when it is chaotic. 
\par 
The response system is synchronized with the drive counterpart when the Lyapunov exponent changes sign. From Fig.$3$ we observe such changes for $\mu$ values ranging from $-1.5$ to $1.25$. In Fig.$4$ we plot $x(t)$ (thin line) and $y(t)$ (thick line), the solutions of the equations for the drive and response systems, as a function of time $t$. This Figure shows that during time evolution $y(t)$ is only partially synchronized with $x(t)$ in the sense that the amplitudes of $y(t)$ and $x(t)$ remain unsynchronized but their phases appear to evolve in synchrony. Synchronization of this type, often called the phase synchronization, occurs when the coupled oscillators are not identical \cite{9}. From $(1)$ and $(8)$, we see that the coupled oscillators of our one-way system are nonidentical. As a result the drive and response systems are found to be partially synchronized.
\section{Conclusion}

In point mechanics and classical field theory there are many physical systems for which the equations of motion do not follow the action principle. One of the nonmaximal way to bring such equations in the framework of variational principle consists in doubling the degrees of freedom of the system. We followed this route to make the Van der Pol equation variationally self-adjoint and thereby construct a one-way coupled system. The system constructed by us involves two non-identical oscillators. Traditionally, the drive-response scenario of chaos is studied by using one-way coupled systems composed of identical oscillators. The system introduced by us may play a role in biology since one-way entrainment occurs in heart. A pacemaker entrains the nerve cells of heart to prevent heart attacks. Such an applicative relevance appears to be further justified by the historically significant work of Van der Pol and Van der Mark who used coupled oscillators discovered by them to make the first model for an artificial pacemaker. 
\section*{Acknowledgements}
This work is supported by the University Grants Commission, Government of India, through grant No. F. 32-39/2006(SR)

\end{document}